\documentclass[sigconf,screen]{acmart}

\setcopyright{acmlicensed}
\acmPrice{15.00}
\acmDOI{10.1145/3395363.3397376}
\acmYear{2020}
\copyrightyear{2020}
\acmSubmissionID{issta20main-id40-p}
\acmISBN{978-1-4503-8008-9/20/07}
\acmConference[ISSTA '20]{Proceedings of the 29th ACM SIGSOFT International Symposium on Software Testing and Analysis}{July 18--22, 2020}{Virtual Event, USA}
\acmBooktitle{Proceedings of the 29th ACM SIGSOFT International Symposium on Software Testing and Analysis (ISSTA '20), July 18--22, 2020, Virtual Event, USA}

\usepackage{graphicx}

\usepackage{dashrule}

\usepackage[linesnumbered,ruled,vlined]{algorithm2e}
\usepackage{algpseudocode}
\usepackage{amsmath,amssymb}
\usepackage{multirow}
\usepackage{xspace}
\usepackage{hhline}

\usepackage{url}
\DeclareFixedFont{\ttb}{T1}{txtt}{bx}{n}{10} 
\DeclareFixedFont{\ttm}{T1}{txtt}{m}{n}{10}  

\usepackage{color}
\definecolor{deepblue}{rgb}{0,0,0.5}
\definecolor{deepred}{rgb}{0.6,0,0}
\definecolor{deepgreen}{rgb}{0,0.5,0}

\usepackage{listings}

\newcommand\pythonstyle{\lstset{
		language=Python,
		basicstyle=\ttm,
		otherkeywords={self},             
		keywordstyle=\ttb\color{deepblue},
		emph={MyClass,__init__},          
		emphstyle=\ttb\color{deepred},    
		stringstyle=\color{deepgreen},
		frame=tb,                         
		showstringspaces=false,            %
		numbers=left
}}

\makeatletter
\def\lst@makecaption{%
  \def\@captype{table}%
  \@makecaption
}
\makeatother

\lstnewenvironment{python}[1][]
{
	\pythonstyle
	\lstset{#1}
}
{}

\usepackage[text size=small, backgroundcolor=blue!30,bordercolor=blue,linecolor=blue]{todonotes}

\newcommand{\substepseparator}{\hspace{1cm}}

\newcommand{\swat}{SWaT\xspace}

\begin{document}

\title{Active Fuzzing for Testing and Securing Cyber-Physical Systems}

\author{Yuqi Chen}
\affiliation{\institution{Singapore Management University}\country{Singapore}}
\email{yuqichen@smu.edu.sg}

\author{Bohan Xuan}
\affiliation{\institution{Zhejiang University}\country{China}}
\email{xuanbohan@zju.edu.cn}

\author{Christopher M. Poskitt}
\orcid{0000-0002-9376-2471}
\affiliation{\institution{Singapore Management University}\country{Singapore}}
\email{cposkitt@smu.edu.sg}

\author{Jun Sun}
\orcid{0000-0002-3545-1392}
\affiliation{\institution{Singapore Management University}\country{Singapore}}
\email{junsun@smu.edu.sg}

\author{Fan Zhang}
\orcid{0000-0001-6087-8243}
\additionalaffiliation{\institution{Zhejiang Lab}, and \institution{Alibaba-Zhejiang University Joint Institute of Frontier Technologies}}
\affiliation{\institution{Zhejiang University}\country{China}}
\email{fanzhang@zju.edu.cn}

\begin{abstract}
	Cyber-physical systems~(CPSs) in critical infrastructure face a pervasive threat from attackers, motivating research into a variety of countermeasures for securing them. Assessing the effectiveness of these countermeasures is challenging, however, as realistic benchmarks of attacks are difficult to manually construct, blindly testing is ineffective due to the enormous search spaces and resource requirements, and intelligent fuzzing approaches require impractical amounts of data and network access. In this work, we propose \emph{active fuzzing}, an automatic approach for finding test suites of packet-level CPS network attacks, targeting scenarios in which attackers can observe sensors and manipulate packets, but have no existing knowledge about the payload encodings. Our approach learns regression models for predicting sensor values that will result from sampled network packets, and uses these predictions to guide a search for payload manipulations (i.e.~bit flips) most likely to drive the CPS into an unsafe state. Key to our solution is the use of \emph{online active learning}, which iteratively updates the models by sampling payloads that are estimated to maximally improve them. We evaluate the efficacy of active fuzzing by implementing it for a water purification plant testbed, finding it can automatically discover a test suite of flow, pressure, and over/underflow attacks, all with substantially less time, data, and network access than the most comparable approach. Finally, we demonstrate that our prediction models can also be utilised as countermeasures themselves, implementing them as anomaly detectors and early warning systems.
\end{abstract}



\maketitle

\section{Introduction}

Cyber-physical systems~(CPSs), characterised by their tight and complex integration of computational and physical processes, are often used in the automation of critical public infrastructure~\cite{US-NSF18a}. Given the potential impact of cyber-attacks on these systems~\cite{Hassanzadeh-et_al19a,Leyden16a,ICS-Cert-Alert16a}, ensuring their security and protection has become a more important goal than ever before. The different temporal scales, modalities, and process interactions in CPSs, however, pose a significant challenge for solutions to overcome, and have led to a variety of research into different possible countermeasures, including ones based on anomaly detection~\cite{Cheng-Tian-Yao17a,Harada-et_al17a,Inoue-et_al17a,Pasqualetti-Dorfler-Bullo11a,Aggarwal-et_al18a,Aoudi-et_al18a,He-et_al19a,Kravchik-Shabtai18a,Lin-et_al18a,Narayanan-Bobba18a,Schneider-Boettinger18a}, fingerprinting~\cite{Ahmed-et_al18a,Ahmed-et_al18b,Gu-et_al18a,Kneib-Huth18a}, invariant-based monitoring~\cite{Cardenas-et_al11a,Adepu-Mathur16a,Adepu-Mathur16b,Chen-Poskitt-Sun16a,Adepu-Mathur18b,Chen-Poskitt-Sun18a,Choi-et_al18a,Giraldo-et_al18a,Yoong-Palleti-Silva-Poskitt20a}, and trusted execution environments~\cite{Spensky-et_al20a}.

Assessing how effective these different countermeasures are at detecting and preventing attacks is another challenge in itself. A typical solution is to use established benchmarks of attacks~\cite{CPS-Datasets,Goh-et_al16a}, which have the advantage of facilitating direct comparisons between approaches, e.g.~as done so in~\cite{Inoue-et_al17a,Kravchik-Shabtai18a,Lin-et_al18a}. Such benchmarks, unfortunately, are few and far between: constructing them manually requires a great deal of time and expertise in the targeted CPS (all while risking insider bias), and generalising them from one CPS to another is a non-starter given the distinct processes, behaviours, and complexities that different systems exhibit.

An alternative solution is to generate benchmarks using \emph{automated testing and fuzzing}, with these techniques overcoming the complexity of CPSs by having access to machine learning~(ML) models trained on their data (e.g.~logs of sensor readings, actuator states, or network traffic). Existing solutions of this kind, however, tend to make unrealistic assumptions about an attacker's capabilities, or require a large body of training data that might not be available. The fuzzer of~\cite{Chen-Poskitt-et_al19a}, for example, can automatically identify actuator configurations that drive the physical states of CPSs to different extremes, but the technology assumes the attacker to have total control of the network and actuators, and is underpinned by a prediction model trained on complete sets of data logs from \emph{several days} of operation. Blindly fuzzing without such a model, however, is ineffective at finding attacks: first, because the search spaces of typical CPSs are \emph{enormous}; and second, because of the wasted time and resources required to be able to observe the effects on a system's physical processes.

In this paper, we present \emph{active fuzzing}, an automatic approach for finding test suites of packet-level CPS network attacks, targeting scenarios in which training data is limited, and in which attackers can observe sensors and manipulate network packets but have no existing knowledge about the encodings of their payloads. Our approach constructs regression models for predicting future sensor readings from network packets, and uses these models to guide a search for payload manipulations that systematically drive the system into unsafe states. To overcome the search space and resource costs, our solution utilises \emph{(online) active learning}~\cite{Lughofer17a}, a form of supervised ML that iteratively re-trains a model on examples that are estimated to maximally improve it. We apply it to CPSs by flipping bits of existing payloads in a way that is guided by one of two frameworks: Expected Model Change Maximization~\cite{Cai-Zhang-Zhou13a}, and a novel adaptation of it based on maximising behaviour change. We query the effects of sampled payloads by spoofing them in the network, updating the model based on the observed effect.

We evaluate our approach by implementing it for the Secure Water Treatment (SWaT) testbed~\cite{SWaT-Reference}, a scaled-down version of a real-world water purification plant, able to produce up to five gallons of drinking water per minute. SWaT is a complex multi-stage system involving chemical processes such as ultrafiltration, de-chlorination, and reverse osmosis. Communication in the testbed is organised into a layered network hierarchy, in which we target the ring networks at the `lowest' level that exchange data using EtherNet/IP over UDP. Our implementation manipulates the binary string payloads of 16 different types of packets, which when considered together have up to $2^{2752}$ different combinations.

Despite the enormous search space, we find that active fuzzing is effective at discovering packet-level flow, pressure, and over/underflow attacks, achieving comparable coverage to an established benchmark~\cite{Goh-et_al16a} and an LSTM-based fuzzer~\cite{Chen-Poskitt-et_al19a} but with substantially less training time, data, and network access. Furthermore, by manipulating the bits of payloads directly, active fuzzing bypasses the logic checks enforced by the system's controllers. These attacks are more sophisticated than those of the LSTM-based fuzzer~\cite{Chen-Poskitt-et_al19a}, which can only generate high-level actuator commands and is unable to manipulate such packets. Finally, we investigate the utility of the learnt models in a different role: defending a CPS directly. We use them to implement anomaly detection and early warning systems for SWaT, finding that when models are suitably expressive, they are effective at detecting both random and known attacks.

\substepseparator

\noindent\textbf{Summary of Contributions.} We present active fuzzing, a black-box approach for automatically discovering packet-level network attacks on real-world CPSs. By iteratively constructing a model with active learning, we demonstrate how to overcome enormous search spaces and resource costs by sampling new examples that maximally improve the model, and propose a new algorithm that guides this process by seeking maximally different behaviour. We evaluate the efficacy of the approach by implementing it for a complex real-world critical infrastructure testbed, and show that it achieves comparable coverage to an established benchmark and LSTM-based fuzzer but with significantly less data, time, and network access. Finally, we show that the learnt models are also effective as anomaly detectors and early warning systems.

\substepseparator

\noindent\textbf{Organisation.} In Section~\ref{sec:background}, we introduce the SWaT testbed, with a particular focus on its network and the structure of its packets. In Section~\ref{sec:active_fuzzing}, we present the components of our active fuzzing approach, and explain how to implement it both in general and for SWaT. In Section~\ref{sec:evaluation}, we evaluate the efficacy of our approach at finding packet-level attacks, and investigate secondary applications of our models as anomaly detectors and early warning systems. In Section~\ref{sec:related_work}, we discuss some related work, then conclude in Section~\ref{sec:conclusion}.

\section{Background}\label{sec:background}

In the following, we present an overview of SWaT, a water treatment testbed that forms the critical infrastructure case study we evaluate active fuzzing on. We describe in more detail its network hierarchy and the structure of its packets, before stating the assumptions we make about the capabilities of attackers.

\substepseparator

\begin{figure}[!t]
	\centering
	\includegraphics[width=0.7\linewidth]{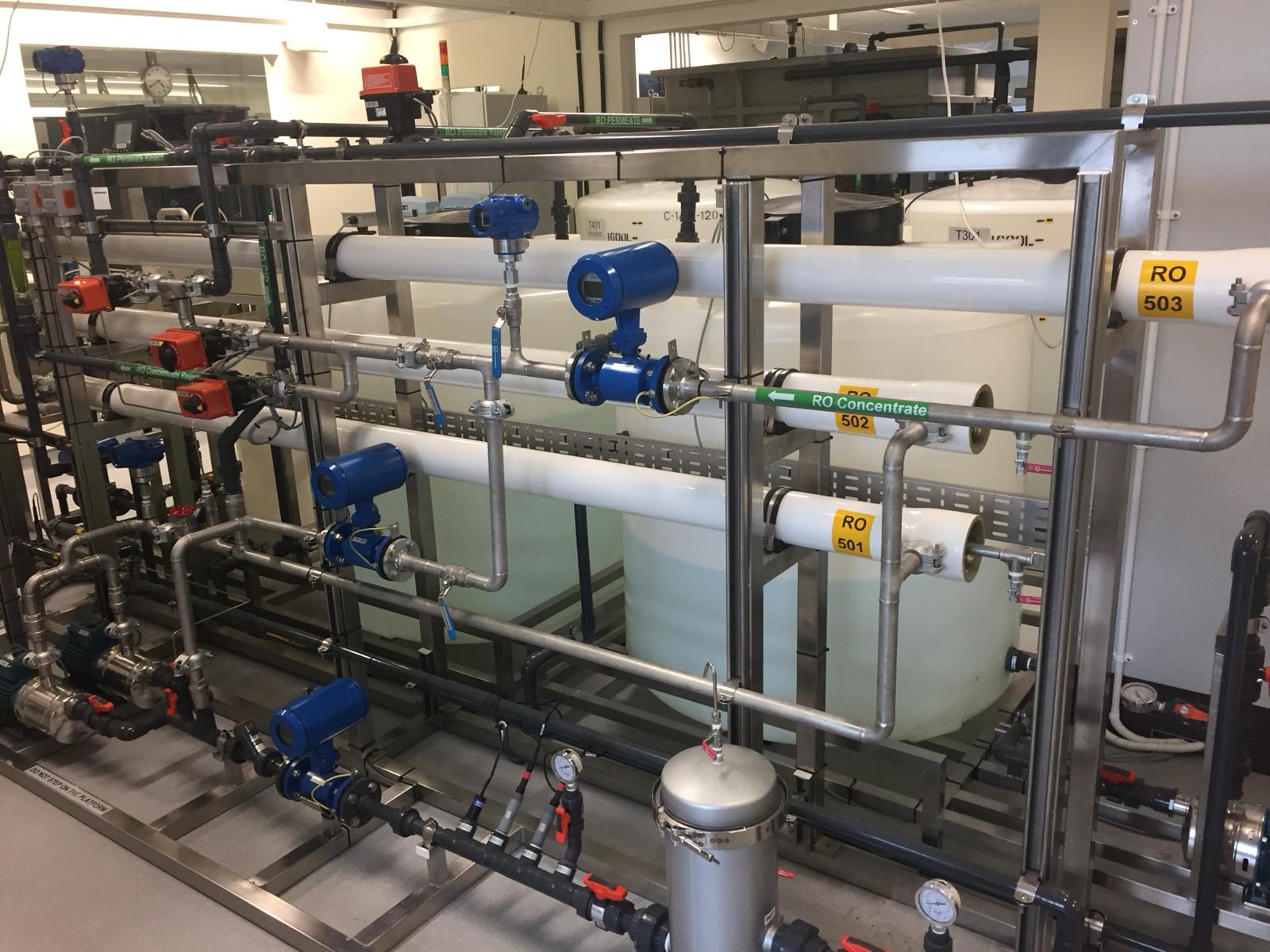}
	\caption{The Secure Water Treatment (SWaT) testbed}
	\label{fig:swat_testbed}
\end{figure}

\noindent\textbf{SWaT Testbed.} The CPS forming the case study of this paper is Secure Water Treatment (SWaT)~\cite{SWaT-Reference}, a scaled-down version of a real-world water purification plant, able to produce up to five gallons of safe drinking water per minute. SWaT (Figure~\ref{fig:swat_testbed}) is intended to be a \emph{testbed} for advancing cyber-security research on critical infrastructure, with the potential for successful technologies to be transferred to the actual plants it is based on. The testbed has been the subject of multiple hackathons~\cite{Adepu-Mathur18a} involving researchers from both academia and industry, and over the years has established a benchmark of attacks to evaluate defence mechanisms against~\cite{Goh-et_al16a}.

SWaT treats water across multiple distinct but co-operating stages, involving a variety of complex chemical processes, such as de-chlorination, reverse osmosis, and ultrafiltration. Each stage in the CPS is controlled by a dedicated Allen-Bradley ControlLogix programmable logic controller~(PLC), which communicates with the sensors and actuators relevant to that stage over a ring network, and with other PLCs over a star network. Each PLC cycles through its program, computing the appropriate commands to send to actuators based on the latest sensor readings received as input. The system consists of 42 sensors and actuators in total, with sensors monitoring physical properties such as tank levels, flow, pressure, and pH values, and actuators including motorised valves (for opening an inflow pipe) and pumps (for emptying a tank). A historian regularly records the sensor readings and actuator commands during SWaT's operation. SCADA software and tools developed by Rockwell Automation are available to support some analyses.

The sensors in SWaT are associated with manufacturer-defined ranges of \emph{safe} values, which in normal operation, they are expected to remain within. If a sensor reports a (true) reading outside of this range, we say the physical state of the CPS has become \emph{unsafe}. If a level indicator transmitter, for example, reports that the tank in stage one has become more than a certain percentage full (or empty), then the physical state has become unsafe due to the risk of an overflow (or underflow). Unsafe pressure states indicate the risk of a pipe bursting, and unsafe levels of water flow indicate the risk of possible cascading effects in other parts of the system.

SWaT implements a number of standard safety and security measures for water treatment plants, such as alarms (reported to the operator) for when these thresholds are crossed, and logic checks for commands that are exchanged between the PLCs. In addition, several attack defence mechanisms developed by researchers have been installed~(see Section~\ref{sec:related_work}).

\begin{figure}[!t]
	\centering\scriptsize
\begin{verbatim}
###[ Ethernet ]### 
  dst       = e4:90:69:a3:0c:f6
  src       = 00:1d:9c:c8:03:e7
  type      = IPv4
###[ IP ]### 
     version   = 4
     ihl       = 5
     tos       = 0xbc
     len       = 68
     id        = 18067
     flags     = 
     frag      = 0
     ttl       = 64
     proto     = udp
     chksum    = 0xb1f3
     src       = 192.168.0.10
     dst       = 192.168.0.12
     \options   \
###[ UDP ]### 
        sport     = 2222
        dport     = 2222
        len       = 48
        chksum    = 0xfca2
###[ ENIP_CPF ]### 
           count     = 2
           \items     \
            |###[ CPF_AddressDataItem ]### 
            |  type_id   = Sequenced Address Item
            |  length    = 8
            |###[ CPF_SequencedAddressItem ]### 
            |     connection_id= 469820023
            |     sequence_number= 18743
            |###[ CPF_AddressDataItem ]### 
            |  type_id   = Connected Transport Packet
            |  length    = 22
            |###[ Raw ]### 
            |     load      = '~2\x01\x00\x00\x00\x00\x00
                               \r\x00\x00\x00\x00\x00\x00
                               \x00\x00\x00\x00\x00\x00
                               \x00'
\end{verbatim}
	\caption{A SWaT packet after dissection by Scapy}
	\label{fig:dissected_packet}
\end{figure}

The network of the SWaT testbed is organised into a layered hierarchy compliant with the ISA99 standard~\cite{ISA99-Standard}, providing different levels of segmentation and traffic control. The `upper' layers of the hierarchy, Levels 3 and 2, respectively handle operation management (e.g.~the historian) and supervisory control (e.g.~touch panel, engineering workstation). Level 1 is a star network connecting the PLCs, and implements the Common Industrial Protocol~(CIP) over EtherNet/IP. Finally, the `lowest' layer of the hierarchy is Level 0, which consists of ring networks (EtherNet/IP over UDP) that connect individual PLCs to their relevant sensors and actuators.

Tools such as Wireshark~\cite{Wireshark} and Scapy~\cite{Scapy} can be used to dissect the header information of a Level 0 SWaT packet, as illustrated in Figure~\ref{fig:dissected_packet}. Here, the source IP address ($\mathtt{192.168.0.10}$) and target IP address ($\mathtt{192.168.0.12}$) correspond respectively to PLC1 and its remote IO device. Actuator commands (e.g.~``open valve MV101'') are encoded in the binary string payloads of these packets. In Figure~\ref{fig:dissected_packet}, the payload is 22 bytes long, but Level 0 packets can also have a payload length of 10 or 32 bytes. Randomly manipulating the payloads has limited use given the size of the search space ($2^{2752}$ possibilities when considering the 16 types of packets we sample; see Section~\ref{sec:packet_sniffing}). Our solution uses active learning to overcome this enormous search space, establishing how different bits impact the physical state without requiring any knowledge of the encoding.

\substepseparator

\noindent\textbf{Attacker Model.} In this work, we assume that attackers have knowledge of the network protocol (e.g. EtherNet/IP over UDP at Level 0 of SWaT), and thus are able to intercept (unencrypted) packets, dissect their header information, and manipulate their payloads. We assume that the packet payloads are binary strings, but \emph{do not} assume any existing knowledge about their meaning or encoding schemes. We assume that attackers can always access the `true' sensor readings while the system is operating, in order to be able to observe the effects of a packet manipulation, or to judge whether or not an attack was successful. These live sensor readings can be observed over several minutes at a time in order to perform some pre-training and active learning, but in contrast to other approaches (e.g.~\cite{Chen-Poskitt-et_al19a}), we do not require access to extensive sets of data for offline learning, and we do not require the ability to arbitrarily issue high-level actuator commands across the system---we do so \emph{only} by manipulating payloads.

\section{Active Fuzzing}\label{sec:active_fuzzing}

Our approach for automatically finding packet-level network attacks in CPSs consists of the following steps. First, data is \emph{collected}: packets are sniffed from the network, their payloads are extracted, and (true) sensor readings are queried. Second, we \emph{pre-train} initial regression models, that take concatenations of packet payloads and predict the future effects on given sensors. Third, we apply an \emph{online active learning} framework, iteratively improving the current model by sampling payloads estimated to maximally improve it. Finally, we search for candidate attacks by flipping important bits in packet payloads, and using our learnt models to identify which of them will drive the system to a targeted unsafe state.

\begin{algorithm}[!t]
\caption{High-Level Overview of Active Fuzzing}\label{alg:overall}
\small
\KwIn{Sensor $s$, prediction time $t_s$, pre-training time $t_p$}
\KwOut{Prediction model $M_s$}
Sniff packets and observe values of $s$ for $t_p$ minutes;\\
Construct a sequence $P$ of feature (bit-)vectors from packet payloads;\\
Construct a sequence $V$ such that each $V[i]$ contains the value of $s$ observed $t_s$ seconds after $P[i]$ was sniffed;\\
(Pre-)train a regression model $M_s$ predicting $V$ from $P$; \\
\Repeat{timeout}
{
	Sample a new feature vector $p$ using an active learning framework (Section~\ref{sec:active_learning_attack_discovery});\\
	Wait for $t_s$ seconds then observe the value $v_s$ of $s$; \\
	$P := P^\frown\langle p\rangle^{t_s}$; [concatenation of $t_s$ copies] \label{alg:line_concat}\\
	$V := V^\frown\langle v_s\rangle^{t_s}$; \\
}
Re-train $M_s$ to predict $V$ from $P$;\\
\Return model $M_s$;
\end{algorithm}

Algorithm~\ref{alg:overall} presents the high-level algorithm of these steps for active fuzzing. Note that the notation in Line~\ref{alg:line_concat} indicates concatenation of sequences. In particular, $t_s$ copies of the vector $p$ are appended to sequence $P$ to add additional weight to the new example when the model is re-trained.

In the following, we describe the steps of the algorithm in more detail, and present the details of one particular implementation for the SWaT water purification testbed.

\subsection{Packet Sniffing and Pre-Training}\label{sec:packet_sniffing}

\noindent\textbf{Collecting Raw Data.} Both the pre-training and active learning phases of our approach require access to two types of data from the CPS under test. First, they must be able to sniff the network packets and extract their binary string payloads. Second, they must be able to access the true readings of any sensors under consideration, as the idea is to be able to observe the effects on sensor readings of different payload manipulations.

For SWaT, our approach extracts packets from Level 0 of the network hierarchy, i.e.~the packets exchanged between PLCs and remote IO devices. By targeting this lowest level of the network, we ensure that our manipulations are altering the actuator states directly. For our prototype, we physically connect some Raspberry Pis to the PLCs of SWaT and sniff the packets using a network connection bridge; in reality, an attacker might sniff packets by other means, e.g.~exploiting the wireless connection when enabled. As Level 0 implements the EtherNet/IP protocol, we can use the tcpdump packet analyser to capture packets, and Scapy to further extract their contents.

For our prototype, we obtain the current sensor readings by querying SWaT's historian. We assume that the historian's data is true, i.e.~that the system is not simultaneously under attack by another entity, and that it is operationally healthy. In reality, an attacker might access this information through an exploit in the historian, e.g.~an EternalBlue exploit~\cite{CVE-2017-0144}, or a backdoor connection (both of which were discovered in SWaT hackathons~\cite{Adepu-Mathur18a}).

\substepseparator

\noindent\textbf{Pre-Training Models.} A goal of our approach is to minimise the amount of data required to train an accurate prediction model for sensor readings. We thus proceed in two phases: a pre-training phase, and an active learning phase. The pre-training phase uses network data to construct an \emph{initial} prediction model, the idea being that it provides a reasonable enough starting point such that active learning will later converge. A key distinction between the two stages is how the attacker behaves: while pre-training, they sit silently to observe \emph{normal} packets of the system; but while actively learning, they intervene by injecting (possibly) \emph{abnormal} packets and then observe the effects. It is thus important to minimise the amount of the time spent in the latter phase to avoid detection.

We require a series of regression models, one per sensor, that take as input the payloads of captured packets, and return as output a prediction of how the considered (true) sensor reading will evolve after a time period. To achieve this goal requires a number of system-specific decisions to be made, for example, the types of packets to train the model on, and a fixed time period that is appropriate to the processes involved (some will change the physical state more quickly than others). There are several types of regression models that are fit for the task. In this work, we focus on two: \emph{linear models} and \emph{gradient-boosting decision trees~(GBDT)}~\cite{Friedman01a}. A linear model is the simplest possible choice and thus serves as our baseline, whereas the GBDT is a well-known and popular example of a non-linear model. Both models can be integrated with existing active learning frameworks for regression, which was a key reason for their selection. Several more expressive models, such as neural networks, do not have any good online active learning strategies (to the best of our knowledge).

In SWaT, packets are collected from Level 0 (see Section~\ref{sec:background}) in the first four stages of the system. By observing the network traffic, we identified four different types of packets in each stage based on payload lengths and headers: packets that have payloads of length (1)~10 bytes; (2)~32 bytes; (3)~22 bytes, with a source IP address of $\mathtt{192.168.0.}S\mathtt{0}$; and (4)~22 bytes, with a source IP address of $\mathtt{192.168.0.}S\mathtt{2}$. Here, $S$ is replaced with the given stage of the system (1, 2, 3, or 4). Across these four stages, there are thus 16 different types of packets in total. In constructing a feature vector for training, we make no assumptions about the meaning of these different packets, so select the first of \emph{each} type of packet that is collected at a particular time point and concatenate their payloads together in a fixed order. This leads to feature vectors containing a series of 2752 bits.

\begin{figure}[!t]
	\centering
	\includegraphics[width=0.95\linewidth]{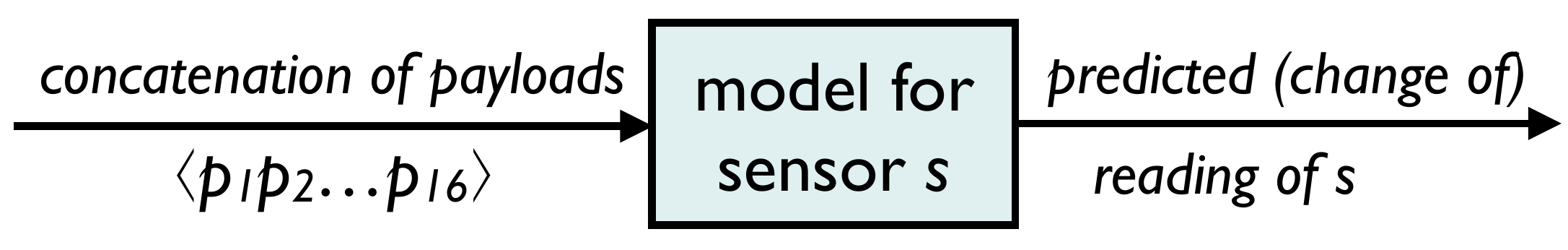}
	\caption{Input/output of a learnt model for sensor $s$}
	\label{fig:model_io}
\end{figure}

Along with constructing a sufficient number of feature vectors (experimentally determined in Section~\ref{sec:evaluation}), we also query the historian for sensor values after fixed time periods have passed. For flow and pressure sensors this time period is 5 seconds; for tank level sensors, it is 30 seconds, owing to the fact that they change state rather more slowly. With this data collected, we train linear and GBDT models for each individual sensor in turn, such that a sensor reading can be predicted for a given bit vector of payloads from the 16 types. An overview of the input/output of these models in given in Figure~\ref{fig:model_io}. Note that for flow and pressure sensors, the corresponding models predict their future \emph{values}, whereas for tank level sensors, the corresponding models predict by how much they will change. This discrepancy is due to the fact that the effects of flow/pressure attacks stabilise at a final value very quickly.

\subsection{Active Learning and Attack Discovery}\label{sec:active_learning_attack_discovery}

\noindent\textbf{Active Learning.} After completing the pre-training phase, we should now have a model that is capable of making some reasonable predictions with respect to \emph{normal} packets in the CPS network. However, the attacks we need for testing the CPS are not necessarily composed of normal packets. We need to train the model further on a broader set of examples, but cannot do it blindly owing to the expense of running the system and the enormity of the search space ($2^{2752}$ potential combinations of feature vectors in SWaT).

Our solution is to train the model further using \emph{(online) active learning}~\cite{Lughofer17a}, a supervised ML approach that iteratively improves the current model. Theoretical studies have shown that active learning may exponentially reduce the amount of training data needed, e.g.~\cite{Fine-et_al02a,Gilad-Bachrach-Navot-Tishby03a,Gilad-Bachrach-Navot-Tishby05a}. The idea is to reduce the amount of additional data by sampling examples that are \emph{estimated} to maximally change the current model in some way. In our case, we use one of two active learning frameworks to guide the construction of new feature vectors by flipping the bits of existing ones (this is more conservative than constructing payloads from scratch, but minimises the possibility of packet rejection). Once new feature vectors have been sampled, we can decompose them into their constituent packets, spoof them in the network, observe their effects on true sensor readings, then re-train the model accordingly.

While active learning for classification problems is well-studied, there are limited active learning frameworks for regression, and some of the ones that exist make assumptions unsuitable for our application (e.g.~a Gaussian distribution~\cite{Castro-Willett-Nowak05a}). However, the Expected Model Change Maximization~(EMCM) approach of Cai et al.~\cite{Cai-Zhang-Zhou13a} avoids this assumption and is suitable for CPSs. Their framework is based on the idea of sampling new examples that are estimated to maximally change the model itself, i.e.~the gradient in linear models, or a linear approximation of GBDTs based on `super features' extracted from the trees (see~\cite{Cai-Zhang-Zhou13a} for details).

Inspired by EMCM, and motivated by the fact we can query live behaviour of the system, we also propose a variant of the framework called Expected Behaviour Change Maximisation~(EBCM). Instead of sampling examples estimated to maximally change the model, EBCM attempts to identify examples that cause maximally different \emph{behaviour} from what the system is currently exhibiting. For example, if a considered sensor reading is increasing, then EBCM may identify examples that cause it to decrease as much as possible instead. The intuition of the approach is that exploring different behaviour in a particular context is more informative. It also seeks to check that unfamiliar packets predicted to cause that behaviour \emph{really do} cause that behaviour, updating the model otherwise.

\begin{algorithm}[!t]
\caption{Expected Behaviour Change Maximisation}\label{alg:EBCM}
\small
\KwIn{Prediction model $M_s$, prediction time $t_s$, maximum number of bits to flip $n_m$}
\KwOut{Feature (bit-)vector $p_f$}
Sniff current packets and construct a feature vector $p_o$ based on their payloads;\\
Wait for $t_s$ seconds then observe the value $v_s$ of $s$;\\
Let $P := \langle\rangle;$ [empty sequence]\\
Let $D := \langle\rangle;$ \\
\Repeat{timeout}
{
	Construct a new vector $p$ from $p_o$ by randomly selecting and flipping $n \leq n_m$ bits;\\
	$v_p := M_s(p)$;\\
	$P := P^\frown\langle p\rangle$;\\
	$D := D^\frown\langle |v_s-v_p|\rangle$;\\
}
Select a feature vector $p_f$ from $P$ using \emph{Roulette Wheel Selection} with corresponding fitness values in $D$;\\
\Return feature (bit-)vector $p_f$;
\end{algorithm}

Algorithm~\ref{alg:EBCM} summarises the steps of EBCM, in which a new feature vector is constructed by sampling additional packets, randomly flipping the bits of several copies, and choosing a vector that would have led to a maximally different reading than the original. Note that to ensure some variation, the feature vector is chosen from a set of several using Roulette Wheel Selection~\cite{Goldberg89a}, which assigns to each candidate a probability of being selected based on its `fitness', here defined as the absolute difference between what the sensor reading actually became (with respect to the original packets) and what the current model predicted for the candidate. If $f_i$ is the fitness of one of $n$ candidates, then its probability of being selected is $f_i / \sum_{j=1}^n f_j$. A random number is generated between 0 and the sum of the candidates' fitness scores. We then iterate through the candidates until the accumulated fitness is larger than that number, returning that final candidate as our chosen bit-vector.

For SWaT, we implemented both EMCM and EBCM, using the same construction of feature vectors (i.e.~a concatenation of the payloads of 16 types of packets). Upon computing new feature vectors using these active learning frameworks, we then break the vectors down into their constituent packets, and spoof them in Level 0 of the network using Scapy. After spoofing, we wait either 5 or 30 seconds (when targeting flow/pressure or tank level sensors respectively) before querying the latest sensor value, then re-train the model based on the new packets and readings observed. This process is repeated until a suitable timeout condition (Section~\ref{sec:experiments_discussion}).

\substepseparator

\noindent\textbf{Attack Discovery and Fuzzing.} In the final step of our approach, we use the learnt models (Figure~\ref{fig:model_io}) to discover attacks, i.e.~packet manipulations that will drive targeted (true) sensor readings out of their safe operational ranges. In particular, after choosing a sensor to target, the corresponding model is used to evaluate a number of candidate packet manipulations and reveal the one that is (predicted) to realise the attack most effectively. The final part of our approach consists of \emph{generating} those candidate packet manipulations for the model to evaluate. 

\begin{algorithm}[!t]
	\caption{Attack Discovery}\label{alg:attack_discovery}
	\small
	\KwIn{Prediction model $M_s$, number of bits to flip $n$, objective function $f$}
	\KwOut{A bit-vector $p_{max}$}
	Sniff current packets and construct a feature vector $p_o$ based on their payloads;\\
	Construct a sequence $\Phi$ of (0-based) indices of $p_o$, from the position with the highest \emph{feature importance} (Section~\ref{sec:active_learning_attack_discovery}) to the lowest;\\
	$k := n-1$; \\
	$Done := \emptyset$; \\
	$f_{max} := 0$; \\
	\Repeat{$k==|\Phi|$ or timeout}
	{
		$\Phi_k := \{i \mid i \in \Phi[0..k] \}$; \\
		$Combs := \{ B \mid B\in 2^{\Phi_k} \wedge |B|=n \wedge B \notin Done \}$; \\
		\For{$c \in Combs$}
		{
			Construct $p$ from $p_o$ by flipping $p_o[i]$ for every $i\in c$;\\
			\If{$f(M_s(p)) > f_{max}$}
			{
				$f_{max} := f(M_s(p))$;\\
				$p_{max} := p$
			}
			$Done := Done \cup c$; \\
		}
		$k := k+1$; \\
	}
	\Return bit-vector $p_{max}$;
\end{algorithm}

Algorithm~\ref{alg:attack_discovery} presents the steps of our packet manipulation procedure for attack discovery. The idea of the algorithm is to identify the bits that are most \emph{important} (i.e.~have the most influence in the prediction), generate candidates by flipping fixed numbers of those bits, before broadening the search to other, less important bits too. As different candidates are generated, they are evaluated against a simple objective function that is maximised as the predicted sensor state becomes closer to an edge of its safe operational range. Suppose that $v_s$ denotes a value of sensor $s$, and that $L_s$ and $H_s$ respectively denote its lower and upper safety thresholds. Let:
\[ d_s = \left\{
        \begin{array}{ll}
            \mathrm{min}\left( \left| v_s - L_s \right|, \left| v_s - H_s \right| \right) & \quad L_s \leq v_s \leq H_s \\
            0 & \quad \mathrm{otherwise}
        \end{array}
    \right. \]

\noindent A suitable objective function that is maximised by values approaching either of the thresholds would then be:
\[ f(v_s) = \frac{1}{d_s / (H_s-L_s)} \]

\noindent We calculate \emph{feature importance} in one of two ways, depending on the model used. For a linear model, the absolute value of the model's weight for that feature is taken as its importance. For a GBDT model, since it is a boosting ensemble model with a bunch of decision trees, we average the feature importance scores of these trees to obtain the feature importance of the overall model.

For SWaT, we implemented attack discovery for multiple flow, pressure, and tank level sensors, and used instances of the objective function above for each of them. The feature vectors returned by Algorithm~\ref{alg:attack_discovery} are broken into their constituent packets, then spoofed in the network using Scapy. If an attack successfully drives a targeted sensor out of its normal operational range (e.g.~over/underflow), we record this, adding the particular packet manipulation used to a test suite of attacks, and document it accordingly (see Section~\ref{sec:evaluation} for an experimental evaluation). Recall that in SWaT, the models for tank levels sensors do not predict future values directly, but rather the magnitude by which they will change by: as a consequence, Algorithm~\ref{alg:attack_discovery} is adapted for these sensors by observing the current reading at the beginning, then using it to calculate the input for the objective function.

\section{Evaluation}\label{sec:evaluation}

We evaluate the effectiveness of active fuzzing for attack discovery and detection using the SWaT testbed~(Section~\ref{sec:background}).

\subsection{Research Questions}

Our evaluation design is centred around the following key research questions~(RQs):

\begin{description}
	\item[\textbf{RQ1 (Training Time):}] How much time is required to learn a high-accuracy model?
	\item[\textbf{RQ2 (Attack Discovery):}] Which model and active learning setup is most effective for attack discovery?
	\item[\textbf{RQ3 (Comparisons):}] How does active fuzzing compare against other CPS fuzzing approaches?
	\item[\textbf{RQ4 (Attack Detection):}] Can the learnt models be used for anomaly detection or early warnings?
\end{description}

\noindent RQ1 is motivated by our assumption that attackers do not have access to large offline datasets for training, and may need to evade anomaly detection systems. How long would an attacker need to spend observing live sensor readings (pre-training) and spoofing packets (active learning) before obtaining a high-accuracy model? RQ2 aims to explore the different combinations of our regression models with and without active learning, in order to establish which is most effective for discovering packet-level CPS attacks, and to quantify any added benefit of active learning in conquering the huge search space. RQ3 is intended to check our work against a baseline, i.e.~its effectiveness in comparison to random search and another guided CPS network fuzzer. Finally, RQ4 aims to explore whether our learnt models can have a secondary application as part of an anomaly detection or early warning system for attacks.

\subsection{Experiments and Discussion}\label{sec:experiments_discussion}

We present the design of our experiments for each of the RQs in turn, as well as some tables of results and the conclusions we draw from them. The programs we developed to implement these experiments on the SWaT testbed are all available online~\cite{Supplementary-Material}.

\substepseparator

\noindent\textbf{RQ1 (Training Time).} Our first RQ aims to assess the amount of time an attacker would require to learn a high-accuracy model from live packets. To answer this question, we design experiments for the two phases of learning in turn.

First, we investigate how long the attacker must spend \emph{pre-training} on normal live sensor readings (i.e.~without any manipulation). Recall that our goal in this phase is not to obtain a highly accurate model, but rather to find a \emph{reasonable} enough model as a starting point for active learning. To do this, we compute the \emph{r2 scores} of linear and GBDT regression models for individual sensors after training for different lengths of time. An r2 score is the percentage of variation explained by the model, and reflects how well correlated the predictions of a sensor and their actual future values are. Prior to training, we collect 230 minutes of packet and sensor data, splitting 180 minutes of it into a training set and the remaining 50 minutes into a test set. For each sensor, we train linear and GBDT models using the full 180 minutes (our upper limit of the experiment), and compute their r2 scores using the test data. We repeat this process for 10 minutes of data, then 20 minutes, \dots\ up to 150 minutes at various intervals until it is clear that model is converging. We judge that a model has converged when the importance scores of its features (see Section~\ref{sec:active_learning_attack_discovery}) have stabilised up to a small tolerance (0.5\% for flow/pressure sensors; 5\% for level sensors) as the model is re-trained on new samples. All steps are repeated ten times and medians are reported to reduce the effects of different starting states.

Second, given a model that has been pre-trained, we investigate how long the attacker must then spend \emph{actively learning} before the model achieves a high accuracy. To do this, we pre-train linear and GBDT prediction models for each sensor for the minimum amount of time previously determined (in the first experiment). Then, for both variants of active learning (Section~\ref{sec:active_learning_attack_discovery}), we sample new sensor data from the system and retrain the models every 5 minutes. In this experiment, we record the amount of time it takes for a model to stabilise with a high r2 score, i.e.~above 0.9, using the same 50 minutes of test data to compute this. We repeat these steps ten times and compute the medians.

\begin{table*}[!t]
	\setlength{\tabcolsep}{7pt}
\caption{r2 scores (\emph{higher is better}) of linear and GBDT sensor prediction models after different amounts of pre-training}
\label{tab:result_pretrain_time}
	\centering\footnotesize
\begin{tabular}{|c|l||cccccccccccc|}
	\cline{3-14}
	\multicolumn{2}{c|}{\multirow{1}{*}{\small\textbf{Linear Models}}} & 10min & 20min & 30min & 40min & 50min & 60min & 70min & 80min & 100min & 120min & 150min & 180min  \\
	\hhline{--============}
	\parbox[t]{2mm}{\multirow{4}{*}{\rotatebox[origin=c]{90}{Flow}}} & FIT101 & 0.3399 & 0.5998 & 0.6698 & 0.7391 & 0.8214 & 0.8475 & {0.8896} & 0.8712 & $\cdots$ & $\cdots$ & $\cdots$ & 0.8840 \\
	
	& FIT201 & -0.7112 & -0.0775 & 0.7394 & 0.8381 & {0.8962} & 0.8931 & $\cdots$ & $\cdots$ & $\cdots$ & $\cdots$ & $\cdots$ & 0.7332 \\
	
	& FIT301 & -0.481 & 0.5292 & 0.8227 & {0.8949} & 0.9121 & 0.9001 & $\cdots$ & $\cdots$ & $\cdots$ & $\cdots$ & $\cdots$ & 0.8772 \\
	
	& FIT401 & -1.2513 & -0.6149 & 0.1123 & 0.3142 & 0.6634 & {0.7235} & 0.6695 & 0.6143 & $\cdots$ & $\cdots$ & $\cdots$ & 0.6425 \\
	
	\hline
	
	Pr. & DPIT301 & -0.2511 & 0.6070 & 0.8648 & {0.9563} & 0.9651 & 0.9642 & $\cdots$ & $\cdots$ & $\cdots$ & $\cdots$ & $\cdots$ & 0.9569 \\
	
	\hline
	
	\parbox[t]{2mm}{\multirow{3}{*}{\rotatebox[origin=c]{90}{T.~Level}}} & LIT101 & 0.0624 & 0.1516 & 0.6024 & 0.6582 & 0.6824 & 0.6168 & 0.7172 & 0.772 & 0.7965 & {0.8197} & 0.8133 & 0.8254 \\
	
	& LIT301 & -0.0806 & 0.0937 & 0.4248 & 0.4949 & 0.5963 & 0.6260 & {0.6583} & 0.4807 & 0.6209 & $\cdots$ & $\cdots$ & 0.5426 \\
	
	& LIT401 & 0.1543 & 0.0612 & 0.2942 & 0.0273 & -0.2208 & 0.1119 & -0.4902 & 0.007 & 0.1259 & 0.2412 & 0.0135 & -0.6597 \\
	 
	 \hline
	\multicolumn{14}{c}{}\\
\end{tabular}

\begin{tabular}{|c|l||cccccccccccc|}
	\cline{3-14}
	\multicolumn{2}{c|}{\multirow{1}{*}{\small\textbf{GBDT Models}}} & 10min & 20min & 30min & 40min & 50min & 60min & 70min & 80min & 100min & 120min & 150min & 180min  \\
	\hhline{--============}
	\parbox[t]{2mm}{\multirow{4}{*}{\rotatebox[origin=c]{90}{Flow}}} & FIT101 & -0.2229 & -0.0245 & 0.4313 & 0.7742 & 0.9112 & 0.9413 & {0.9755} & 0.9741 & $\cdots$ & $\cdots$ & $\cdots$ & 0.9637 \\
	
	& FIT201 & -0.7116 & 0.3545 & {0.9584} & 0.9633 & 0.9504 & 0.9642 & $\cdots$ & $\cdots$ & $\cdots$ & $\cdots$ & $\cdots$ & 0.9421 \\
	
	& FIT301 & -0.9453 & 0.2496 & 0.8051 & {0.9734} & 0.9731 & 0.9751 & $\cdots$ & $\cdots$ & $\cdots$ & $\cdots$ & $\cdots$ & 0.9524 \\
	
	& FIT401 & -0.2124 & 0.3120 & 0.7015 & 0.8125 & 0.8342 & {0.8861} & 0.8453 & 0.7921 & $\cdots$ & $\cdots$ & $\cdots$ & 0.8025 \\
	
	\hline
	
	Pr. & DPIT301 & -0.5508 & 0.8218 & 0.9387 & {0.9757} & 0.9818 & 0.9831 & 0.9912 & 0.9901 & 0.9875 & 0.9706 & 0.9496 & 0.9085 \\
	
	\hline
	
	\parbox[t]{2mm}{\multirow{3}{*}{\rotatebox[origin=c]{90}{T.~Level}}} & LIT101 & -0.042 & -0.1352 & -0.153 & 0.3858 & 0.6459 & 0.7881 & 0.8536 & {0.8680} & 0.8961 & 0.9221 & 0.8721 & 0.9019 \\
	
	& LIT301 & -0.0419 & -0.2151 & -0.0185 & 0.3863 & 0.7059 & {0.8208} & 0.6486 & 0.7938 & 0.5498 & 0.7363 & $\cdots$ & 0.7291 \\
	
	& LIT401 & -1.1415 & 0.1121 & 0.7123 & 0.8377 & 0.8503 & {0.8575} & 0.7731 & 0.7907 & 0.8764 & 0.8743 & 0.8741 & 0.8444 \\
	 
	 \hline
\end{tabular}

\end{table*}

\emph{Results.} Table~\ref{tab:result_pretrain_time} presents the results of our first experiment. The columns correspond to the amount of training time (10 minutes through to 180), whereas the rows correspond to regression models for individual SWaT sensors, including Flow Indicator Transmitters (e.g.~FIT101), a Differential Pressure Indicator Transmitter (DPIT301), and Level Indicator Sensors (e.g.~LIT101). For the LITs, our models predict their values 30 seconds into the future (as tank levels rise very slowly), whereas for all other sensors our models make predictions for 5 seconds ahead. The values reported in the table are r2 scores: here, a score of 1 indicates that the model and test data are perfectly correlated, whereas a score of -1 would indicate that there is no correlation at all. When there is clear evidence of a model converging, we do not repeat the experiment for longer training periods (except 180 minutes, our upper limit).

All of our models eventually converge during pre-training, except the linear model for LIT401: the process involving this tank is too complicated to be represented as a linear model due to the multiple interactions and dependencies involving other stages of the testbed (the GBDT model does not suffer this problem). Note that while pre-training leads to relatively high r2 scores for a number of the models (e.g.~the simpler processes involving flow), this does not necessarily imply that the models will be effective for attack discovery (as we investigate in RQ2). For the goal of determining a minimum amount of pre-training time, we fix it at 40 minutes, as all models (except Linear-LIT401) exhibit some positive correlation by then ($\geq\!0.3$). Some scores are still low, but this will allow us to assess whether active learning is still effective when applied to cases that lack a good pre-trained model.

\begin{table}[!t]
\caption{Median time (mins; \emph{lower is better}) for active learning~(AL) configurations to achieve an r2 score above 0.9}
\label{tab:result_active_learning_time}
	\centering\footnotesize
\begin{tabular}{|l||cccc|c|ccc|}
	\cline{2-9}
	\multicolumn{1}{c|}{} &\multicolumn{4}{c|}{Flow} &\multicolumn{1}{c|}{Pressure} &\multicolumn{3}{c|}{Tank Level}  \\
	\cline{2-9}
	\multicolumn{1}{c|}{\textbf{AL Config.}} &  \rotatebox{90}{FIT101} & \rotatebox{90}{FIT201} & \rotatebox{90}{FIT301} & \rotatebox{90}{FIT401} & \rotatebox{90}{\scriptsize DPIT301} & \rotatebox{90}{LIT101} & \rotatebox{90}{LIT301} & \rotatebox{90}{LIT401}  \\
	\hhline{-========}
	Linear (EBCM) & 25 & 15 & 30 & 15 & 30 & --- & --- & --- \\

	Linear (EMCM) & 20 & 20 & 45 & 15 & 40 & --- & --- & --- \\
	 
	GBDT (EBCM) & 10 & 10 & 25 & 10 & 30 & 35 & 30 & 45 \\
	 
	GBDT (EMCM) & 10 & 10 & 25 & 10 & 20 & 40 & 40 & 45  \\
	 
	 \hline
\end{tabular}
\end{table}

Table~\ref{tab:result_active_learning_time} presents the results of our second experiment. Here, the columns contain the sensors that each regression model is targeting, whereas the rows contain the type of model and active learning variant considered. The values reported are the number of minutes (accurate up to 5 minutes) of active learning that it takes before models achieve an r2 score above 0.9. Note that with active learning, none of the linear models for tank level sensors were able to exceed our r2 threshold (although they did converge for LIT101 and LIT301 with lower scores). All GBDT models were able to exceed the r2 threshold with active learning, indicating that the additional expressiveness is important for some processes of SWaT---likely because the actual processes \emph{are} non-linear. The amount of time required varied from 10 up to 45 minutes. Taking the pre-training time into consideration:

\begin{center}
\noindent\fbox{%
    \parbox{0.75\linewidth}{%
        \small\emph{Once pre-trained on 40 minutes of data observations, attackers can accurately predict SWaT's sensor readings after 10--45 minutes of active learning.}
    }
}
\end{center}

\noindent This is a significantly reduced amount of time compared to SWaT's LSTM-based fuzzer~\cite{Chen-Poskitt-et_al19a}, the model of which was trained for approximately two days on a rich dataset compiled from four days of constant operation.

\substepseparator

\noindent\textbf{RQ2 (Attack Discovery).} Our second RQ aims to assess which combinations of models and active learning setups (including no active learning at all) are most effective for \emph{finding attacks}, i.e.~manipulations of packet payloads that would cause the true readings of a particular sensor to eventually cross one of its safety thresholds (e.g.~risk of overflow, or risk of bursting pipe).

To do this, we experimentally calculate the \emph{success rates} at finding attacks for all variants of models covering the flow, pressure, and tank level sensors. Furthermore, we do so while restricting the manipulation of the packets' payloads to different quantities of bit flips, from 1--5 and 10 such flips. For each model variant, we calculate the success rate by running our active fuzzer 1000 times with the given model, and recording as a percentage the number of times in which the resulting modified packet would cause the physical state to cross\footnote{With the exception of the low threshold for flow sensors, which is $0$.} a safety threshold. Note that it is important to flip existing payload bits, rather than craft packets directly, as the system's built-in validation procedures may reject them.

\begin{table*}[!t]
	\setlength{\tabcolsep}{4.5pt}
\caption{Success rates (\%s; \emph{higher is better}) of different model configurations for finding packet manipulations (1-5 and 10 bit flips) that successfully drive SWaT's flow, pressure, and tank level readings to safety thresholds}
\label{tab:result_bit_flips_part1}

\parbox{.49\linewidth}{
	\flushleft\footnotesize
\begin{tabular}{|c|l||cccc|c|ccc|}
	\cline{3-10}
	\multicolumn{2}{c|}{\multirow{2}{*}{}} &\multicolumn{4}{c|}{Flow} &\multicolumn{1}{c|}{Pr.} &\multicolumn{3}{c|}{Level}  \\
	\cline{3-10}
	\multicolumn{2}{c|}{\textbf{Models (1 Bit Flip)}} &  \rotatebox{90}{FIT101} & \rotatebox{90}{FIT201} & \rotatebox{90}{FIT301} & \rotatebox{90}{FIT401} & \rotatebox{90}{\scriptsize DPIT301} & \rotatebox{90}{LIT101} & \rotatebox{90}{LIT301} & \rotatebox{90}{LIT401}  \\
	\hhline{--========}
	\parbox[t]{2mm}{\multirow{4}{*}{\rotatebox[origin=c]{90}{Linear}}} & Pre-Train Only (40min) & 0.5 & 4.4 & 1.3 & 1.8 & 0.9 & 0 & 0 & 0 \\

	 & Pre-Train Only (90min) & 0.8 & 3.8 & 1.4 & 0.2 & 2.5 & 0 & 0 & 0 \\
	 
	 & Active Learning (EBCM) & 27 & 22.3 & 7.5 & 43.9 & 14.4 & 0 & 0 & 0 \\
	 
	 & Active Learning (EMCM) & 29.4 & 19.2 & 7.9 & 36.6 & 9.1 & 0 & 0 & 0 \\
	 
	 \hline
	
	\parbox[t]{2mm}{\multirow{4}{*}{\rotatebox[origin=c]{90}{GBDT}}} & Pre-Train Only (40min) & 0 & 59.3 & 0 & 0 & 0 & 0 & 0 & 0 \\

	 & Pre-Train Only (90min) & 0 & 57.7 & 30.3 & 0 & 0 & 0 & 0 & 0 \\
	 
	 & Active Learning (EBCM) & 97.7 & 99.2 & 97.9 & 75.4 & 96.1 & 0 & 0 & 0 \\
	 
	 & Active Learning (EMCM) & 97.6 & 99.4 & 97.6 & 13.5 & 95.3 & 0 & 0 & 0 \\

	\hline
	--- & Random (No Model) & 0 & 0 & 0 & 0.2 & 0 & 0 & 0 & 0 \\
	\hline
\end{tabular}
}
\hfill
\parbox{.49\linewidth}{
	\flushright\footnotesize
\begin{tabular}{|c|l||cccc|c|ccc|}
	\cline{3-10}
	\multicolumn{2}{c|}{\multirow{2}{*}{}} &\multicolumn{4}{c|}{Flow} &\multicolumn{1}{c|}{Pr.} &\multicolumn{3}{c|}{Level}  \\
	\cline{3-10}
	\multicolumn{2}{c|}{\textbf{Models (2 Bit Flips)}} &  \rotatebox{90}{FIT101} & \rotatebox{90}{FIT201} & \rotatebox{90}{FIT301} & \rotatebox{90}{FIT401} & \rotatebox{90}{\scriptsize DPIT301} & \rotatebox{90}{LIT101} & \rotatebox{90}{LIT301} & \rotatebox{90}{LIT401}  \\
	\hhline{--========}
	\parbox[t]{2mm}{\multirow{4}{*}{\rotatebox[origin=c]{90}{Linear}}} & Pre-Train Only (40min) & 0.7 & 8 & 1.9 & 3.1 & 1.7 & 0.2 & 0 & 0 \\

	 & Pre-Train Only (90min) & 2.7 & 9.4 & 4.1 & 0.6 & 6.6 & 0 & 0 & 0 \\
	 
	 & Act.~Learning (EBCM) & 46.1 & 39.1 & 15.1 & 77.6 & 30.1 & 2 & 0 & 0 \\
	 
	 & Act.~Learning (EMCM) & 47.4 & 31.8 & 16.5 & 72.4 & 17.9 & 0.6 & 0 & 0 \\
	 
	 \hline
	
	\parbox[t]{2mm}{\multirow{4}{*}{\rotatebox[origin=c]{90}{GBDT}}} & Pre-Train Only (40min) & 0 & 98.7 & 0 & 0 & 0 & 0 & 0 & 0 \\

	 & Pre-Train Only (90min) & 0.3 & 96 & 59 & 0 & 0 & 0 & 0 & 0 \\
	 
	 & Act.~Learning (EBCM) & 100 & 99.9 & 100 & 100 & 99.9 & 76.4 & 0 & 0 \\
	 
	 & Act.~Learning (EMCM) & 100 & 100 & 99.7 & 100 & 99.9 & 87.1 & 0 & 0 \\

	\hline
	--- & Random (No Model) & 0.3 & 0 & 0 & 0.2 & 0 & 0 & 0 & 0 \\
	\hline
\end{tabular}
}

\vspace{7pt}

\setlength{\tabcolsep}{4pt}

\parbox{.49\linewidth}{
	\flushleft\footnotesize
\begin{tabular}{|c|l||cccc|c|ccc|}
	\cline{3-10}
	\multicolumn{2}{c|}{\multirow{2}{*}{}} &\multicolumn{4}{c|}{Flow} &\multicolumn{1}{c|}{Pr.} &\multicolumn{3}{c|}{Level}  \\
	\cline{3-10}
	\multicolumn{2}{c|}{\textbf{Models (3 Bit Flips)}} &  \rotatebox{90}{FIT101} & \rotatebox{90}{FIT201} & \rotatebox{90}{FIT301} & \rotatebox{90}{FIT401} & \rotatebox{90}{\scriptsize DPIT301} & \rotatebox{90}{LIT101} & \rotatebox{90}{LIT301} & \rotatebox{90}{LIT401}  \\
	\hhline{--========}
	\parbox[t]{2mm}{\multirow{4}{*}{\rotatebox[origin=c]{90}{Linear}}} & Pre-Train Only (40min) & 1.2 & 10.6 & 3.5 & 3.5 & 3 & 0 & 0 & 0 \\

	 & Pre-Train Only (90min) & 4.2 & 12.5 & 5.5 & 0.5 & 8.2 & 0 & 0 & 0 \\
	 
	 & Act.~Learning (EBCM) & 58.6 & 50.6 & 25.7 & 91.7 & 37.2 & 4.2 & 0.1 & 0.1 \\
	 
	 & Act.~Learning (EMCM) & 60.4 & 45.1 & 21.4 & 88.7 & 24.2 & 2.7 & 0.3 & 0 \\
	 
	 \hline
	
	\parbox[t]{2mm}{\multirow{4}{*}{\rotatebox[origin=c]{90}{GBDT}}} & Pre-Train Only (40min) & 0 & 100 & 0.1 & 0 & 0 & 0 & 0 & 0 \\

	 & Pre-Train Only (90min) & 1.1 & 100 & 80.6 & 0 & 0 & 0 & 0 & 0 \\
	 
	 & Act.~Learning (EBCM) & 100 & 100 & 100 & 100 & 100 & 97 & 8.1 & 23.7 \\
	 
	 & Act.~Learning (EMCM) & 100 & 100 & 99.7 & 100 & 100 & 99.2 & 3.7 & 32.3 \\

	\hline
	--- & Random (No Model) & 0.1 & 0.2 & 0.1 & 0.1 & 0.3 & 0 & 0 & 0 \\
	\hline
\end{tabular}
	
}
\hfill
\parbox{.49\linewidth}{
	\flushright\footnotesize
\begin{tabular}{|c|l||cccc|c|ccc|}
	\cline{3-10}
	\multicolumn{2}{c|}{\multirow{2}{*}{}} &\multicolumn{4}{c|}{Flow} &\multicolumn{1}{c|}{Pr.} &\multicolumn{3}{c|}{Level}  \\
	\cline{3-10}
	\multicolumn{2}{c|}{\textbf{Models (4 Bit Flips)}} &  \rotatebox{90}{FIT101} & \rotatebox{90}{FIT201} & \rotatebox{90}{FIT301} & \rotatebox{90}{FIT401} & \rotatebox{90}{\scriptsize DPIT301} & \rotatebox{90}{LIT101} & \rotatebox{90}{LIT301} & \rotatebox{90}{LIT401}  \\
	\hhline{--========}
	\parbox[t]{2mm}{\multirow{4}{*}{\rotatebox[origin=c]{90}{Linear}}} & Pre-Train Only (40min) & 0.9 & 13.8 & 5.4 & 4.9 & 4.3 & 0 & 0 & 0 \\

	 & Pre-Train Only (90min) & 6.4 & 15.1 & 7.5 & 0.9 & 12.1 & 0 & 0 & 0 \\
	 
	 & Act.~Learning (EBCM) & 71.6 & 65 & 27.9 & 97.7 & 49.1 & 11.4 & 0.1 & 0.2 \\
	 
	 & Act.~Learning (EMCM) & 75.1 & 26.8 & 31.1 & 95.7 & 29.2 & 5.4 & 0.3 & 0 \\
	 
	 \hline
	
	\parbox[t]{2mm}{\multirow{4}{*}{\rotatebox[origin=c]{90}{GBDT}}} & Pre-Train Only (40min) & 0 & 100 & 0.7 & 0 & 0 & 0 & 0 & 0 \\

	 & Pre-Train Only (90min) & 1.2 & 100 & 96.1 & 0 & 0 & 0 & 0 & 0 \\
	 
	 & Act.~Learning (EBCM) & 100 & 100 & 100 & 100 & 100 & 100 & 20.4 & 55.7 \\
	 
	 & Act.~Learning (EMCM) & 100 & 100 & 99.7 & 100 & 100 & 100 & 17.7 & 67.3 \\

	\hline
	--- & Random (No Model) & 0.1 & 0.1 & 0.1 & 0.2 & 0.2 & 0 & 0 & 0 \\
	\hline
\end{tabular}
}

\vspace{7pt}

\setlength{\tabcolsep}{4pt}

\parbox{.49\linewidth}{
	\flushleft\footnotesize
\begin{tabular}{|c|l||cccc|c|ccc|}
	\cline{3-10}
	\multicolumn{2}{c|}{\multirow{2}{*}{}} &\multicolumn{4}{c|}{Flow} &\multicolumn{1}{c|}{Pr.} &\multicolumn{3}{c|}{Level}  \\
	\cline{3-10}
	\multicolumn{2}{c|}{\textbf{Models (5 Bit Flips)}} &  \rotatebox{90}{FIT101} & \rotatebox{90}{FIT201} & \rotatebox{90}{FIT301} & \rotatebox{90}{FIT401} & \rotatebox{90}{\scriptsize DPIT301} & \rotatebox{90}{LIT101} & \rotatebox{90}{LIT301} & \rotatebox{90}{LIT401}  \\
	\hhline{--========}
	\parbox[t]{2mm}{\multirow{4}{*}{\rotatebox[origin=c]{90}{Linear}}} & Pre-Train Only (40min) & 1.7 & 19.3 & 6.2 & 5.7 & 4.3 & 0.4 & 0 & 0 \\

	 & Pre-Train Only (90min) & 6.6 & 18.4 & 8.7 & 1.1 & 13.5 & 0.1 & 0 & 0 \\
	 
	 & Act.~Learning (EBCM) & 78.3 & 71.6 & 35.4 & 99.5 & 56.6 & 16.3 & 0.5 & 0 \\
	 
	 & Act.~Learning (EMCM) & 81.5 & 62.6 & 36 & 97.9 & 39.3 & 9.3 & 0.5 & 0 \\
	 
	 \hline
	
	\parbox[t]{2mm}{\multirow{4}{*}{\rotatebox[origin=c]{90}{GBDT}}} & Pre-Train Only (40min) & 0 & 100 & 3.1 & 0 & 0 & 0 & 0 & 0 \\

	 & Pre-Train Only (90min) & 3.1 & 100 & 99.8 & 0 & 0.1 & 0 & 0 & 0 \\
	 
	 & Act.~Learning (EBCM) & 100 & 100 & 100 & 100 & 100 & 100 & 29.7 & 76 \\
	 
	 & Act.~Learning (EMCM) & 100 & 100 & 99.7 & 100 & 100 & 100 & 34.9 & 82.2 \\

	\hline
	--- & Random (No Model) & 0.2 & 0.4 & 0.1 & 0.1 & 0.2 & 0 & 0 & 0 \\
	\hline
\end{tabular}
}
\hfill
\parbox{.49\linewidth}{
	\flushright\footnotesize
\begin{tabular}{|c|l||cccc|c|ccc|}
	\cline{3-10}
	\multicolumn{2}{c|}{\multirow{2}{*}{}} &\multicolumn{4}{c|}{Flow} &\multicolumn{1}{c|}{Pr.} &\multicolumn{3}{c|}{Level}  \\
	\cline{3-10}
	\multicolumn{2}{c|}{\textbf{Models (10 Bit Flips)}} &  \rotatebox{90}{FIT101} & \rotatebox{90}{FIT201} & \rotatebox{90}{FIT301} & \rotatebox{90}{FIT401} & \rotatebox{90}{\scriptsize DPIT301} & \rotatebox{90}{LIT101} & \rotatebox{90}{LIT301} & \rotatebox{90}{LIT401}  \\
	\hhline{--========}
	\parbox[t]{2mm}{\multirow{4}{*}{\rotatebox[origin=c]{90}{Linear}}} & Pre-Train Only (40min) & 4.7 & 34.9 & 9.6 & 12.5 & 10.7 & 0.8 & 0 & 0 \\

	 & Pre-Train Only (90min) & 13 & 38.3 & 19.2 & 2.3 & 29.1 & 1 & 0 & 0 \\
	 
	 & Act.~Learning (EBCM) & 96.4 & 91 & 61.8 & 100 & 81.5 & 35.7 & 6.1 & 2.1 \\
	 
	 & Act.~Learning (EMCM) & 96.7 & 89.1 & 59.5 & 100 & 63 & 29.1 & 5.7 & 0 \\
	 
	 \hline
	
	\parbox[t]{2mm}{\multirow{4}{*}{\rotatebox[origin=c]{90}{GBDT}}} & Pre-Train Only (40min) & 0 & 100 & 31.5 & 0 & 0 & 0 & 0 & 0 \\

	 & Pre-Train Only (90min) & 12.1 & 100 & 99.8 & 0 & 2.3 & 0 & 0 & 0 \\
	 
	 & Act.~Learning (EBCM) & 100 & 100 & 100 & 100 & 100 & 100 & 72.2 & 99.3 \\
	 
	 & Act.~Learning (EMCM) & 100 & 100 & 99.7 & 100 & 100 & 100 & 77 & 99.7 \\

	\hline
	--- & Random (No Model) & 0.2 & 0.4 & 0.4 & 0.1 & 0.5 & 0 & 0 & 0 \\
	\hline
\end{tabular}
}

\end{table*}

\emph{Results.} Table~\ref{tab:result_bit_flips_part1} presents the results of our experiment for RQ2. Each sub-table reports on a restriction to a particular number of payload bit flips, ranging from 1--5 and then 10. The columns contain the sensors we are attempting to drive into unsafe states, whereas the rows contain the type of model and active learning variant considered (if any). The final row, Random (No Model), is discussed as part of RQ3. The values recorded are success rates (\%s), where 100\% indicates that all 1000 model-guided bit flips would succeed, and where 0\% indicates that none of them would do. In the active learning models, pre-training was conducted for 40 minutes (as determined in RQ1). We also include a model that was pre-trained \emph{only} for 90 minutes---roughly the time to do \emph{both} pre-training and active learning---to ensure a fair comparison.

We can draw a number of conclusions from these results. First, linear models are not expressive enough in general for driving the bit flipping of payloads: their success rates for the LITs, for example, is mostly 0\%, and even at 10 bit flips numbers for most sensors remain very low. GBDT quite consistently outperforms the linear models, often approaching 100\% success rates. Like the linear models, GBDT struggled to attack the LITs for small numbers of bit flips (likely because multiple commands are needed to affect these sensors), but can attack them all once the restriction is lifted to ten bit flips.

\begin{center}
\noindent\fbox{%
    \parbox{0.75\linewidth}{%
        \small\emph{The expressiveness of the underlying model is critically important: active learning alone is not enough to compensate for this.}
    }
}
\end{center}

Another conclusion that can be drawn from the tables relates to the significantly higher success rates for variants using active learning, both for linear models and GBDT models. The combination of active learning with the expressiveness of GBDT, in particular, leads to attacks being found in all cases for the 10 bit flip restriction. With active learning enabled, the difference is often significant (e.g.~0\% vs.~100\% for FIT401, 10 bit flips). The results suggest that active learning is key for finding the `critical bits' in payloads, given its ability to sample and query new data. Models that have only been pre-trained just recognise trends observed in normal data, and do not necessarily know which bits involved in the patterns are the critical ones for enacting an attack.

\begin{center}
\noindent\fbox{%
    \parbox{0.75\linewidth}{%
        \small\emph{Active learning is effective at identifying critical bits in payloads, and can lead to significantly higher success rates in attack discovery.}
    }
}
\end{center}

\substepseparator

\noindent\textbf{RQ3 (Comparisons).} Our third RQ assesses how active fuzzing performs against two baselines. First, for every sensor (as targeted in RQ2), we randomly generate 1000 $k$-bit payload manipulations (where $k$ is 1--5 or 10) and assess for them the attack success rates (a percentage, as calculated in RQ2). Second, we qualitatively compare our attacks against the ones identified by the LSTM-based fuzzer for SWaT~\cite{Chen-Poskitt-et_al19a} as well as an established benchmark of SWaT network attacks~\cite{Goh-et_al16a} that was manually crafted by experts.

\emph{Results.} The results of the random flipping baseline are given in the final rows of Table~\ref{tab:result_bit_flips_part1}. Clearly, this is not an effective strategy for finding attacks based on packet manipulation, as no success rate exceeds $0.5\%$. This is unsurprising due to the huge search space involved. Note also that for the more challenging over/underflow attacks, random bit flipping is unable to find any examples at all.

Regarding the LSTM-based fuzzer for SWaT~\cite{Chen-Poskitt-et_al19a}, a side-by-side comparison is difficult to make as it does not manipulate packets but rather only issues high-level actuator commands (e.g.``OPEN MV101''). Our approach is able to find attacks spanning the same range of sensed properties, but does so by manipulating the bits of packets directly (closer to the likely behaviour of a real attacker) and without the same level of network control (other than true sensor readings, which both approaches require). In this sense our attacks are more elaborate than those of the LSTM fuzzer. Our approach is also substantially faster: active fuzzing can train effective models in 50-85 minutes, whereas the underlying model used in~\cite{Chen-Poskitt-et_al19a} required approximately two whole days.

Our coverage of the SWaT benchmark~\cite{Goh-et_al16a} is comparable to that of~\cite{Chen-Poskitt-et_al19a}, since both approaches find attacks spanning the same sensed properties. However, all of the attacks in~\cite{Goh-et_al16a} and~\cite{Chen-Poskitt-et_al19a} are implemented at Level 1 of the network. Active fuzzing instead generates packet-manipulating attacks at Level 0, which has the advantage of avoiding interactions with the PLC code, possibly making manipulations harder to detect (e.g.~bypassing command validation checks). In this sense, the attacks that active fuzzing finds complement and enrich the benchmark.

\begin{center}
\noindent\fbox{%
    \parbox{0.75\linewidth}{%
        \small\emph{Active fuzzing finds attacks covering the same sensors as comparable work, but with significantly less training time, and by manipulating packets directly.}
    }
}
\end{center}

\substepseparator

\noindent\textbf{RQ4 (Attack Detection).} Our final RQ considers whether our learnt models can be used not only for attack discovery but also attack \emph{prevention}. In particular, we investigate their use in two defence mechanisms: an anomaly detector and an early warning system. We then assess how effective they are detecting attacks.

\begin{table}[!t]
	\setlength{\tabcolsep}{2pt}
\caption{Success rates (\%s; \emph{higher is better}) of different anomaly detector models at detecting injected sensor values}
\label{tab:result_anomaly_detector}
	\centering\footnotesize
\begin{tabular}{|c|l||cccc|c|ccc|}
	\cline{3-10}
	\multicolumn{2}{c|}{\multirow{2}{*}{}} &\multicolumn{4}{c|}{Flow} &\multicolumn{1}{c|}{Pr.} &\multicolumn{3}{c|}{Tank Level}  \\
	\cline{3-10}
	\multicolumn{2}{c|}{\small\textbf{Anomaly Detector}} &  \rotatebox{90}{FIT101} & \rotatebox{90}{FIT201} & \rotatebox{90}{FIT301} & \rotatebox{90}{FIT401} & \rotatebox{90}{\scriptsize DPIT301} & \rotatebox{90}{LIT101} & \rotatebox{90}{LIT301} & \rotatebox{90}{LIT401}  \\
	\hhline{--========}
	\parbox[t]{2mm}{\multirow{4}{*}{\rotatebox[origin=c]{90}{Linear}}} & Pre-Train Only (40min) & 82.3 & 100 & 100 & 41.3$^\ast$ & 100 & 8.2$^\ast$ & 38.5$^\ast$ & 40.1$^\ast$ \\

	 & Pre-Train Only (90min) & 100 & 100 & 100 & 62.9$^\ast$ & 100 & 67.8$^\ast$ & 39.3$^\ast$ & 13.2$^\ast$ \\
	 
	 & Act.~Learning (EBCM) & 100 & 100 & 100 & 71.1$^\ast$ & 100 & 59.8$^\ast$ & 51.2$^\ast$ & 64.1$^\ast$ \\
	 
	 & Act.~Learning (EMCM) & 100 & 100 & 100 & 71.4$^\ast$ & 100 & 57.5$^\ast$ & 44.7$^\ast$ & 50.7$^\ast$ \\
	 
	 \hline
	
	\parbox[t]{2mm}{\multirow{4}{*}{\rotatebox[origin=c]{90}{GBDT}}} & Pre-Train Only (40min) & 100 & 100 & 100 & 100 & 100 & 71.8$^\ast$ & 74.6$^\ast$ & 76$^\ast$ \\

	 & Pre-Train Only (90min) & 100 & 100 & 100 & 100 & 100 & 95.3 & 92.3 & 74.1 \\
	 
	 & Act.~Learning (EBCM) & 100 & 100 & 100 & 100 & 100 & 91.8 & 95.5 & 84.1 \\
	 
	 & Act.~Learning (EMCM) & 100 & 100 & 100 & 100 & 100 & 92.5 & 95.8 & 83.4 \\

	\hline
\end{tabular}
\end{table}

To perform anomaly detection, we continuously perform the following process: we read the current values of sensors, and then use our learnt models to predict their values 5 seconds into the future (30 seconds for tank levels). After 5 (or 30) seconds have passed, the \emph{actual} values $v_a$ are compared with those that were predicted $v_p$, and an anomaly is reported if $|v_p-v_a|/v_m>0.05$ (or $|v_p-v_a|>5$ for tanks), where $v_m$ is the largest possible observable value for the sensor. To evaluate the effectiveness of this detection scheme, we implement an experiment on actual historian data extracted from SWaT~\cite{Goh-et_al16a}. For each sensor in turn, we randomly generate 1000 spoofed sensor values by randomly adding or subtracting values (in the range 5-10 for LITs, or $0.05v_m$ through to $0.1v_m$ for the others) to sensor readings at different points of the data. We then use our learnt models to determine what would have been predicted from the data 5 or 30 seconds earlier, comparing the actual and predicted values as described. We record the success rates of our anomaly detectors at detecting these spoofed sensor readings.

Our early warning system is set up in a similar way, continuously predicting the future readings of sensors based on the current network traffic. The key difference is that rather than comparing actual values with previously predicted values, we instead issue warnings at the time of prediction if the future value of a sensor is outside of its well-defined normal operational range. To experimentally assess this, we manually subject the SWaT testbed to the Level 1 attacks identified in~\cite{Chen-Poskitt-et_al19a} (which itself covers more unsafe states that the SWaT benchmark~\cite{Goh-et_al16a}), targeting each sensor in turn. When each attack is underway, we use our learnt models to predict the future sensor readings. If a warning is issued at some point \emph{before} a sensor is driven outside of its normal range, we record this as a success. We repeat this ten times for each sensor.

\emph{Results.} Table~\ref{tab:result_anomaly_detector} contains the results of our anomaly detection experiment. The columns indicate the sensors for which values in the data were manipulated, whereas the rows indicate the model and active learning variant used. The values are the success rates, i.e.~the percentage of spoofed sensor values that were detected as anomalous. Asterisks ($\ast$) indicate where false positive rates were above $5\%$, meaning the anomaly detectors were not practically useful. For the flow and pressure sensors, most variants of model and active learning were able to successfully detect anomalies, the main exception being FIT401 for which the linear model performed poorly. The tank level sensors were more challenging to perform anomaly detection for, but the GBDT models have a clear edge over the linear ones. Active learning made little difference across the experiments, except to improve the accuracy of the original 40 minute pre-trained models.

\begin{table}[!t]
\caption{Success rates (\%s; \emph{higher is better}) of different models at warning before sensors exit their safe ranges}
\label{tab:result_warning_system}
	\centering\footnotesize
\begin{tabular}{|c|l||ccc|}
	\cline{3-5}
	\multicolumn{2}{c|}{\multirow{2}{*}{}} &\multicolumn{3}{c|}{Tank Level}  \\
	\cline{3-5}
	\multicolumn{2}{c|}{\textbf{Early Warning System}} &  LIT101 & LIT301 & LIT401  \\
	\hhline{--===}
	\multicolumn{2}{|c||}{Linear (all variants)} & --- & --- & --- \\
	 
	 \hline
	
	\parbox[t]{2mm}{\multirow{4}{*}{\rotatebox[origin=c]{90}{GBDT}}} & Pre-Train Only (40min) & --- & --- & --- \\

	 & Pre-Train Only (90min) & 100 & 100 & 100 \\
	 
	 & Active Learning (EBCM) & 100 & 100 & 100 \\
	 
	 & Active Learning (EMCM) & 100 & 100 & 100 \\

	\hline
\end{tabular}
\end{table}

Table~\ref{tab:result_warning_system} contains the results of our early warning detection experiment. The columns indicate the sensed properties that were targeted by attacks (e.g.~drive LIT101 outside of its safe range), whereas the rows indicate the model and active learning variant used. The values are the success rates, i.e.~the percentages of attacks that were warned about before succeeding. Cells containing dashes (---) indicate that more than 5\% of the warnings were false positive, and thus too unreliable. The first thing to note is that the experiment only considered the tank level sensors: this is because the flow and pressure sensors can be forced into unsafe states very quickly, requiring more immediate measures than an early warning system. The tanks however take time to fill up or empty, and thus are a more meaningful target for this solution. Second, the model has a clear impact: GBDT models with either active learning or at least 90 minutes of pre-training are accurate enough to warn about 100\% of the attacks, whereas the linear models are not expressive enough and suffer from false positives. Again, active learning improves the accuracy of 40 minute pre-trained models sufficiently, but otherwise is not critical: its key role is not in prevention but in \emph{discovering} attacks, through its ability to identify the critical bits to manipulate.

\begin{center}
\noindent\fbox{%
    \parbox{0.75\linewidth}{%
        \small\emph{Our models can be repurposed as anomaly detectors or early warning systems, but active learning is not as critical here as in attack discovery.}
    }
}
\end{center}

\subsection{Threats to Validity}

While our work has been extensively evaluated on a real critical infrastructure testbed, threats to the validity of our conclusions of course remain. First, while SWaT is a fully operational water treatment testbed, it is not as large as the plants it is based on, meaning our results may not scale-up (this is difficult to assess, as access to such plants is subject to strict confidentially). Second, it may not generalise to CPSs in domains that have different operational characteristics, motivating future work to properly assess this. Finally, while our anomaly detector performed well, our sensor spoofing attacks were generated randomly, and may not be representative of a real attacker's behaviour (note however that the early warning system was assessed using previously documented attacks). Similarly, our early warning detection systems performed well at detecting known over/underflow attacks, but these attacks are of the kind that active fuzzing itself can generate: how the models perform against different kinds of attacks requires further investigation.

\section{Related Work}\label{sec:related_work}

In this section, we highlight a selection of the literature that is related to the main themes of this paper: \emph{learning from traffic} (including active learning), \emph{defending CPSs}, and \emph{testing/verifying CPSs}.

\substepseparator

\noindent\textbf{Learning from Network Traffic.} The application of machine learning to network traffic is a vibrant area of research~\cite{Nguyen-Armitage08a}, but models are typically constructed to perform \emph{classification} tasks. To highlight a few examples: Zhang et al.~\cite{Zhang-et_al15a} combine supervised and unsupervised ML to learn models that can classify zero-day traffic; Nguyen and Armitage~\cite{Nguyen-Armitage06a} learn from statistical features of sub-flows to classify between regular consumer traffic and traffic originating from online games; and Atkinson et al.~\cite{Atkinson-et_al18a} use a classifier to infer personal information by analysing encrypted traffic patterns caused by mobile app usage. All these examples are in contrast to active fuzzing, where \emph{regression} models are learnt for predicting how a set of network packets will cause a (true) sensor reading of a CPS to change. We are not aware of other work building regression models in a similar context.

Similar to active fuzzing, there are some works that apply active learning, but again for the purpose of classification, rather than regression. Morgan~\cite{Morgan15a}, for example, uses active learning to reduce training time for streaming data classifiers, as do Zhao and Hoi~\cite{Zhao-Hoi} but for malicious URL classifiers.

\substepseparator

\noindent\textbf{Defending CPSs.} Several different research directions on detecting and preventing CPS attacks have emerged in the last few years. Popular approaches include anomaly detection, where data logs (e.g.~from historians) are analysed for suspicious events or patterns~\cite{Cheng-Tian-Yao17a,Harada-et_al17a,Inoue-et_al17a,Pasqualetti-Dorfler-Bullo11a,Aggarwal-et_al18a,Aoudi-et_al18a,He-et_al19a,Kravchik-Shabtai18a,Lin-et_al18a,Narayanan-Bobba18a,Schneider-Boettinger18a}; digital fingerprinting, where sensors are checked for spoofing by monitoring time and frequency domain features from sensor and process noise~\cite{Ahmed-et_al18a,Ahmed-et_al18b,Gu-et_al18a,Kneib-Huth18a}; and invariant-based checks, where conditions over processes and components are constantly monitored~\cite{Cardenas-et_al11a,Adepu-Mathur16a,Adepu-Mathur16b,Chen-Poskitt-Sun16a,Adepu-Mathur18b,Chen-Poskitt-Sun18a,Choi-et_al18a,Giraldo-et_al18a}. These techniques are meant to complement and go beyond the built-in validation procedures installed in CPSs, which typically focus on simpler and more localised properties of the system.

The strengths and weaknesses of different countermeasures has been the focus of various studies. Urbina et al.~\cite{Urbina-et_al16a} evaluated several attack detection mechanisms in a comprehensive review, concluding that many of them are not limiting the impact of stealthy attacks (i.e.~from attackers who have knowledge about the system's defences), and suggest ways of mitigating this. C\'{a}rdenas et al.~\cite{Cardenas-et_al14a} propose a general framework for assessing attack detection mechanisms, but in contrast to the previous works, focus on the business cases between different solutions. For example, they consider the cost-benefit trade-offs and attack threats associated with different methods, e.g.~centralised vs.~distributed.

As a testbed dedicated for cyber-security research, many different countermeasures have been developed for SWaT itself. These include anomaly detectors, typically trained on the publicly released dataset~\cite{Goh-et_al16a,CPS-Datasets} using unsupervised learning techniques, e.g.~\cite{Inoue-et_al17a,Kravchik-Shabtai18a,Goh_et-al17a}. A supervised learning approach is pursued by~\cite{Chen-Poskitt-Sun16a,Chen-Poskitt-Sun18a}, who inject faults into the PLC code of (a high-fidelity simulator) in order to obtain abnormal data for training. Ahmed et al.~\cite{Ahmed-et_al18a,Ahmed-et_al18b} implemented fingerprinting systems based on sensor and process noise for detecting spoofing. Adepu and Mathur~\cite{Adepu-Mathur16a,Adepu-Mathur16b,Adepu-Mathur18b} systematically and manually derived physics-based invariants and other conditions to be monitored during the operation of SWaT. Feng et al.~\cite{Feng-et_al19a} also generate invariants, but use an approach based on learning and data mining that can capture noise in sensor measurements more easily than manual approaches.

\substepseparator
 
\noindent\textbf{Testing and Verifying CPSs.} Several authors have sought to improve the defences of CPSs by constructing or synthesising attacks that demonstrate flaws to be fixed. Liu et al.~\cite{Liu-et_al11a} and Huang et al.~\cite{Huang-et_al18a}, for example, synthesise attacks for power grids that can bypass bad measurement detection systems and other conventional monitors. Dash et al.~\cite{Dash-Karimibiuki-Pattabiraman19a} target robotic vehicles, which are typically protected using control-based monitors, and demonstrate three types of stealthy attacks that evade detection. Uluagac et al.~\cite{Uluagac-et_al14a} presented attacks on sensory channels (e.g.~light, infrared, acoustic, and seismic), and used them to inform the design of an intrusion detection system for sensory channel threats. Active fuzzing shares this goal of identifying attacks in order to improve CPS defences.

\emph{Fuzzing} is a popular technique for automatically testing the defences of systems, by providing them with invalid, unexpected, or random input and monitoring how they respond. Our active fuzzing approach does exactly this, guiding the construction of input (network packets) using prediction models, and then observing sensor readings to understand how the system responds. The closest fuzzing work to ours is~\cite{Chen-Poskitt-et_al19a}, which uses an LSTM-based model to generate actuator configurations, but requires vast amounts of data and system access to function effectively. Fuzzing has also been applied for testing CPS models, e.g.~CyFuzz~\cite{Chowdhury-Johnson-Csallner17a} and DeepFuzzSL~\cite{Shrestha-Chowdhury-Csallner20a}, which target models developed in Simulink. Outside of the CPS domain, several fuzzing tools are available for software: American fuzzy lop~\cite{Zalewski}, for example, uses genetic algorithms to increase the code coverage of tests; Cha et al.~\cite{Cha-Woo-Brumley15a} use white-box symbolic analysis on execution traces to maximise the number of bugs they find; and grammar-based fuzzers (e.g.~\cite{Holler-Herzig-Zeller12a,Godefroid-Peleg-Singh17a}) use formal grammars to generate complex structured input, such as HTML/JavaScript for testing web browsers. Fuzzing can also be applied to network protocols in order to test their intrusion detection systems (e.g.~\cite{Vigna-et_al04a}). Our work, in contrast, assumes that an attacker has \emph{already compromised} the network (as per Section~\ref{sec:background}).

There are techniques beyond fuzzing available for analysing CPS models in Simulink. A number of authors have proposed automated approaches for falsifying such models, i.e.~for finding counterexamples of formal properties. To achieve this, Yamagata et al.~\cite{Akazaki-et_al18a,Yamagata-et_al20a} use deep reinforcement learning, and Silvetti et al.~\cite{Silvetti-Policriti-Bortolussi17a} use active learning. Chen et al.~\cite{Chen-Sabato-Kong16a} also use active learning, but for mining formal requirements from CPS models. Note that unlike these approaches, active fuzzing is applied directly at the network packet level of a real and complex CPS, and therefore does not make any of the abstractions that modelling languages necessitate.

A number of approaches exist that allow for CPSs to be \emph{formally} verified or analysed. These typically require a formal specification or model, which, if available in the first place, may abstract away important complexities of full-fledged CPS processes. Kang et al.~\cite{Kang-et_al16a}, for example, construct a discretised first-order model of \swat's first three stages in Alloy, and analyse it with respect to some safety properties. This work, however, uses high-level abstractions of the physical process, only partially models the system, and would not generalise to the packet-level analyses that active fuzzing performs. Sugumar and Mathur~\cite{Sugumar-Mathur17a} analyse CPSs using timed automata models, simulating their behaviour under single-stage single-point attacks. Castellanos et al.~\cite{Castellanos-Ochoa-Zhou18a}, McLaughlin et al.~\cite{McLaughlin-et_al14a}, and Zhang et al.~\cite{Zhang-et_al19a} perform formal analyses based on models extracted from the PLC programs, whereas Etigowni et al.~\cite{Etigowni-et_al16a} analyse information flow using symbolic execution. If a CPS can be modelled as a hybrid system, then a number of formal techniques may be applied, including model checking~\cite{Frehse-et_al11a,Wang-et_al18a}, SMT solving~\cite{Gao-Kong-Clarke13a}, reachability analysis~\cite{Johnson-et_al16a}, non-standard analysis~\cite{Hasuo-Suenaga12a}, process calculi~\cite{Lanotte-et_al17a}, concolic testing~\cite{Kong-et_al16a}, and theorem proving~\cite{Quesel-et_al16a}. Defining a formal model that accurately characterises enough of the CPS, however, is the \emph{hardest} part, especially for techniques such as active fuzzing that operate directly at the level of packet payloads.

\section{Conclusion}\label{sec:conclusion}

We proposed \emph{active fuzzing}, a black-box approach for automatically building test suites of packet-level CPS network attacks, overcoming the enormous search spaces and resource costs of such systems. Our approach learnt regression models for predicting future sensor values from the binary string payloads of network packets, and used these models to identify payload manipulations that would achieve specific attack goals (i.e.~pushing true sensor values outside of their safe operational ranges). Key to achieving this was our use of online active learning, which reduced the amount of training data needed by sampling examples that were estimated to maximally improve the model. We adapted the EMCM~\cite{Cai-Zhang-Zhou13a} active learning framework to CPSs, and proposed a new version of it that guided the process by maximising behaviour change.

We presented algorithms for implementing active fuzzing, but also demonstrated its efficacy by implementing it for the SWaT testbed, a multi-stage water purification plant involving complex physical and chemical processes. Our approach was able to achieve comparable coverage to an established benchmark and LSTM-based fuzzer, but with significantly less data, training time, and resource usage. Furthermore, this coverage was achieved by more sophisticated attacks than those of the LSTM-based fuzzer, which can only generate high-level actuator commands and is unable to manipulate packets directly. Finally, we showed that the models constructed in active learning were not only useful for attack \emph{discovery}, but also for attack \emph{detection}, by implementing them as anomaly detectors and early warning systems for SWaT. We subjected the plant to a series of random sensor-modification attacks as well as existing actuator-manipulation attacks, finding that our most expressive learnt models were effective at detecting them.

\begin{acks}
	We are grateful to the three anonymous ISSTA referees for their very constructive feedback. This research / project is supported by the National Research Foundation, Singapore, under its National Satellite of Excellence Programme ``Design Science and Technology for Secure Critical Infrastructure'' (Award Number:~NSoE\_DeST-SCI2019-0008). Any opinions, findings and conclusions or recommendations expressed in this material are those of the author(s) and do not reflect the views of National Research Foundation, Singapore. It is also supported in part by a Major Scientific Research Project of Zhejiang Lab (2018FD0ZX01), Alibaba-Zhejiang University Joint Institute of Frontier Technologies, Zhejiang Key R\&D Plan (2019C03133), the Fundamental Research Funds for the Central Universities (2020QNA5021), and by an SUTD-ZJU IDEA Grant for Visiting Professor (SUTD-ZJU VP 201901).
\end{acks}

\bibliographystyle{ACM-Reference-Format}
\balance
\bibliography{references}

\end{document}